\documentclass{camera}
\usepackage{graphicx,amsmath,amssymb,bm}

\newcommand{\calM}{{\cal M}}
\newcommand{\calT}{{\cal T}}
\newcommand{\mred}{m_{\rm red}}

\begin{document}

%%%%%%%%%%%%%%%%%%%%%%%%%%%%%%%%%%%%%%%%%%%%%%%%%%%%%%%%
% The title, only the first letter capitalized; if you want to split it in
% two or more lines, put a \\ macro at each line break
% example: 
%   \title{Title: first line\\ second line}
%
\title{Emission of neutrino-antineutrino pairs by hadronic
bremsstrahlung processes}

%%%%%%%%%%%%%%%%%%%%%%%%%%%%%%%%%%%%%%%%%%%%%%%%%%%%%%%%
% The author(s), separated by commas; do not put a
% comma before the last author, use instead the \and
% macro which produces a normal ``and'' in the
% caps/small caps context
%
\author{Sonia Bacca$^{1,2}$, Rishi Sharma$^3$ \and Achim Schwenk$^{4,5}$}

%%%%%%%%%%%%%%%%%%%%%%%%%%%%%%%%%%%%%%%%%%%%%%%%%%%%%%%%
%
\organization{$^1$ TRIUMF, Vancouver,  British Columbia, V6T 2A3, Canada \\
$^2$ Department of Physics and Astronomy, University of Manitoba, Winnipeg, MB, R3T 2N2, Canada\\
$^3$ TIFR, Homi Bhabha Road, Navy Nagar, Mumbai 400005, India \\
$^4$ Institut f\"{u}r Kernphysik, Technische Universit\"{a}t Darmstadt, 64289 Darmstadt, Germany\\
$^5$ ExtreMe Matter Institute EMMI, GSI Helmholtzzentrum f\"{u}r Schwerionenforschung GmbH,
64291 Darmstadt, Germany}

\maketitle

\begin{abstract}
We review our recent calculations of neutrino-antineutrino pair production from bremsstrahlung processes
in hadronic collisions at temperature and densities relevant for core-collapse supernovae. We focus  on
neutron-neutron and neutron-$\alpha$ collisions.
\end{abstract}

%%%%%%%%%%%%%%%%%%%%%%%%%%%%%%%%%%%%%%%%%%%%%%%%%%%%%%%%
% Write the text starting from here and using the usual
% LaTeX commands.
%
\section{Introduction}
Understanding core-collapse supernovae and neutron star formation requires input from nuclear physics regarding not only the equation of state, but also transport processes.
In core-collapse supernovae, neutrinos of all flavors are liberated seconds after the collapse
and take away 99$\%$ of the total gravitational energy. These are dominantly produced in the proto-neutron star
created after the collapse and propagate in the dense material around the
neutrinosphere while undergoing interactions with themselves and with 
protons, neutrons and nuclei in this material \cite{Janka12, Burrows13} before eventually
decoupling. Interactions of neutrinos with this material affect their spectrum
and  play a role in the dynamics of the shock wave passing through it.

Consequently, the identification and quantitative determination of the
processes that lead to neutrino production, scattering and absorption for the
relevant astrophysical conditions is crucial. Here, we concentrate on the
calculation of the neutrino-antineutrino pair production and absorption by
bremsstrahlung from hadronic collisions.  This process is induced by the
neutral current interaction and is the only hardonic process that produces muon and
tau neutrinos, which unlike electron neutrinos are not produced in
charged-current reactions like $\beta$-decays and electron captures.

Supernova explosions are most sensitive to neutrino processes near the surface
of the proto-neutron star, where matter is very neutron-rich, and at densities
of the order of one tenth of the nuclear saturation density $\rho_0 = 2.8
\times 10^{14}$ g cm$ ^{−3}$.  The bremsstrahlung process involving a
neutron-neutron collision is the dominant mechanism for production of muon and
tau neutrinos in this region and involves nuclear interactions. While this
process has been relatively well studied (see Refs.~\cite{Bacca09,Bacca12} and references therein),
the question about the relevance of neutrino production from other hadronic
collisions involving neutrons and protons \cite{Bartl14} and neutrons and finite nuclei~\cite{Sharma15}
has only been explored recently. Below we review our recent calculations \cite{Bacca09,Bacca12, Sharma15} of the neutrino-antineutrino pair production from bremsstrahlung
processes and compare results from neutron-neutron \cite{Bacca09,Bacca12} and
neutron-$\alpha$ collisions~\cite{Sharma15}.

\section{Results}
\begin{figure}
\begin{centering}
\includegraphics[width=4cm]{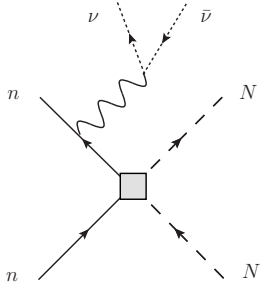}
\caption{Feynman diagram for the neutrino-antineutrino (dotted lines) bremsstrahlung process from hadronic collisions: solid lines are neutrons, while dashed lines are either neutrons or $\alpha$-particles and the wiggly line is the $Z^0$ boson. The square box represents the hadronic interactions taking place in the collision.}
\label{fig1} % optional figure label, must be unique
\end{centering}
\end{figure}

We consider the $\nu\bar{\nu}$ bremsstrahlung from neutron ($n$) collisions with  other neutrons or $\alpha$-particles (denoted as $N$). The Feynman diagram of this process is shown in Fig.~\ref{fig1}. The weak vertex radiating the $Z^0$ boson can come either from the incoming neutron or the outgoing neutron. Moreover, in case that $N$ in Fig.~\ref{fig1} is another neutron,  two additional
diagrams associated with the $Z^0$ boson emitted from the right in-going or outgoing neutron exist (see Ref.~\cite{Bacca09, Sharma15} for more details).

In case of $n\alpha$ collisions, the scattering amplitude $\calM$ for this process can be
written as
\begin{equation}
i \calM = -\frac{i \, G_F C_A}{\sqrt{2}} \frac{1}{\omega} 
\sum_{j=1,2,3} l^j
 \chi_1^\dagger \, [{\boldsymbol \sigma}^j , \calT({\bf k}) ] \,
\chi_3 \,, \label{eq:generalM}
\end{equation}
where $G_F$ is the Fermi coupling constant and $C_A=-g_A/2$ is
the axial-vector coupling for neutrons. Here, $\omega$ is the energy transferred from the neutrino-pair to the
neutron,  ${\bf k}$ is the momentum
transfer,
$l^j$ is the leptonic current, ${\boldsymbol \sigma}^j$ are Pauli
matrices associated with the neutron spin, $\chi_{1,3}$ are neutron
spinors, and $\calT({\bf k})$ denotes the strong interaction scattering vertex. 
From the fact that the commutator of the neutron spin operator with
the strong vertex appears in Eq.~(\ref{eq:generalM}), one can readily see that only noncentral
terms of the strong force will contribute to the amplitude. Similarly, also for $nn$ collisions only noncentral 
terms of the strong force contribute.

\begin{figure}
\begin{centering}
  \includegraphics[width=9cm]{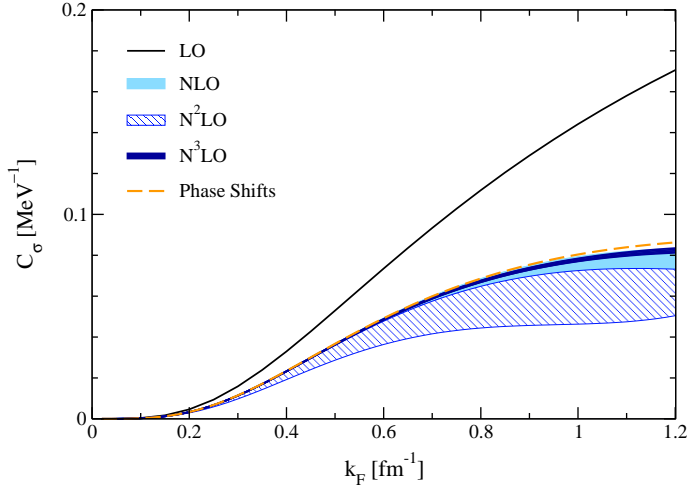}
  \caption{(Color online) Spin relaxation rate in $nn$ collisions as a function of Fermi momentum $k_F$ obtained from chiral EFT
interactions at different orders. The leading order corresponds to the one-pion-exchange potential. The calculation obtained from nucleon-nucleon phase shifts~\cite{phaseshifts} is also shown for comparison.}
  \label{nn1}
\end{centering}
\end{figure}

For $nn$ bremsstrahlung, this process was studied first by Friman and Maxwell~\cite{Friman79} using the Born approximation  to calculate the scattering vertex $\calT$ with the nucleon-nucleon interaction described by the one-pion-exchange  potential, which gives rise to the long-range tensor force. In Ref.~\cite{Hanhart01}, it was calculated for degenerate conditions using nucleon-nucleon phase shifts. We extended these studies in Ref.~\cite{Bacca09}  by using chiral effective field theory (EFT) interactions at different orders ~\cite{Epelbaum06, Entem03} and solving the Boltzmann equation with collisions in a relaxation time approximation.
 Our main result is shown in Fig.~\ref{nn1}, where we give the spin relaxation rate, a quantity proportional to the bremsstrahlung rate to a good approximation (see Ref.~\cite{Bacca09} for more details), as a function of the Fermi momentum $k_F$ of the nucleons in degenerate conditions. Note that $k_F=1.2$ fm$^{-1}$  corresponds to a density of about $1 \times 10^{14}$ g cm$^{-3}$. It is worth remarking that, given that central contact terms do not contribute to this process, the leading order (LO) in EFT corresponds to the case of the one-pion-exchange potential, typically used in supernovae simulations~\cite{HR}. Calculations at next-to-next-to-next-to-leading order (N$^3$LO) differ by about a factor of 2-3 from the LO results.   The largest decrease of the strength is observed in going from the LO to next-to-leading order (NLO), where two-pion-exchange contributions introduce further tensor forces.

In Fig.~\ref{nn1} we also show the comparison of our  N$^3$LO calculation with the result obtained using the Nijmegen partial-wave analysis of experimental nucleon-nucleon scattering data~\cite{phaseshifts}. The fact that the latter is very similar to the Born approximation of the  N$^3$LO potentials indicate that the involved $nn$ amplitudes are perturbative. 

Motivated by  nondegenerate conditions encountered in supernovae,
we have also investigated this regime~\cite{Bacca12}, essentially describing nucleons with  Boltzmann distributions instead of Fermi distributions. We have observed also in this case that
 N$^3$LO calculations are lower than LO results by a factor of 2-3.

While for  systems consisting of neutrons only, the noncentral part of nuclear forces give non-zero rates, 
in the case of proton admixture also the central part contributes, because the axial charge of the neutron and proton are unequal. Thus, also LO contact terms  and  $S$-wave scattering are important. This was recently investigated in Ref.~\cite{Bartl14}, where it was observed that using the $\calT$ matrix from phase shifts leads to different results than N$^3$LO at low densities, indicating that central interactions are nonperturbative.  Moreover,
relative to the one-pion-exchange potential, the $\cal T$ matrix leads to enhanced neutrino rates at low densities and reduced rates  at higher densities, which could lead to interesting mechanisms for energy transport in the outer layers of the supernova, due to  
 competing neutrino processes at different temperatures and densities.

After nucleon-nucleon collisions have been addressed, a natural follow-up question is: What is the effect of
neutrino-antineutrino pair production from collisions of nucleons with finite nuclei?
 To tackle this question,  we recently have addressed the case of neutron-$\alpha$ collisions~\cite{Sharma15}. Some $\alpha$ particles ($^4$He nuclei) are in fact present in the outer layers of the proto-neutron star~\cite{Janka07, arcones}.
\begin{figure}
\begin{centering}
  \includegraphics[width=9cm]{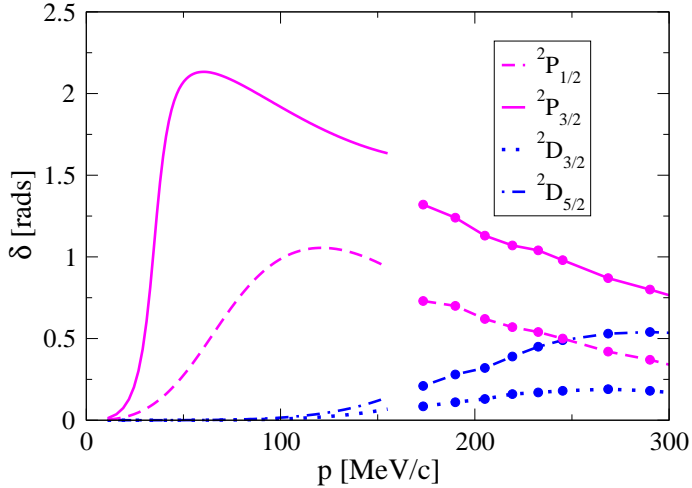}
  \caption{(Color online) Neutron-$\alpha$ scattering phase shifts $\delta$ as a function of the relative momentum $p$. The different partial waves are
labeled by the standard notation $^{2S+1}\ell_j$ . Phase shifts are shown
following Ref.~\cite{Horowitz06}: for low momenta from the fits to data of Ref.~\cite{Arndt70}
and for higher momenta (lines with points) from optical model
calculations \cite{Amos}. $S$-waves are not shown because they do not contribute to the bremsstrahlung process.}
  \label{delta}
\end{centering}
\end{figure}
 While the density of
$\alpha$ particles is smaller than the density of neutrons, given that
$n\alpha$ scattering features a resonance at low energies in the $P$-waves as shown in Fig.~\ref{delta}, one expects
the bremsstrahlung rate per particle pair from this process to be enhanced
due to a larger ${\cal{T}}(k)$, as is clear from Eq.~(\ref{eq:generalM}).
The $P$-wave resonance in the $n\alpha$ system can be seen as a single particle excitation on top of the  filled $s$-shell of the $\alpha$ particle.
The spin-orbit interaction splits the $^2P_{3/2}$ from the $^2P_{1/2}$ channel, hence the $P$-waves are expected to contribute to the bremsstrahlung process. 

\begin{figure}
\begin{centering}
  \includegraphics[width=8cm]{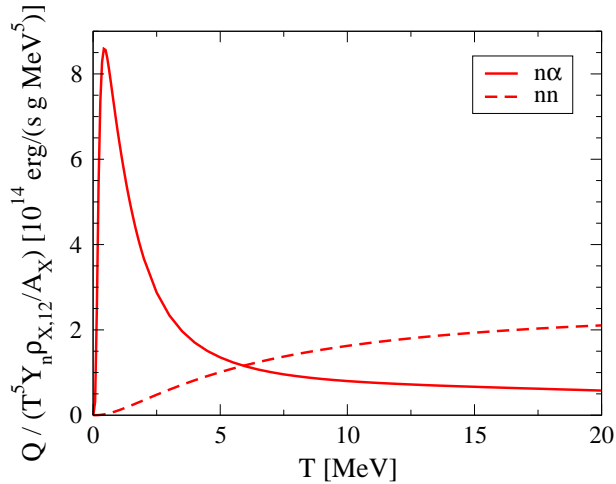}
  \caption{(Color online)  Energy-loss rates as a function of
temperature T. The solid (dashed) line is for $n$-$\alpha$ ($n$-$n$) bremsstrahlung.}
  \label{emiss}
\end{centering}
\end{figure}

As for the $nn$ case, 
we estimate the rate of the $n\alpha$ process by evaluating the $\calT$ matrix starting from the experimental phase shifts shown in Fig.~\ref{delta} as 
\begin{equation}
\langle p \ell S j| {\cal T} | p \ell S j \rangle =
(2\pi)^3 \frac{p^2}{\pi \, \mred} \frac{1- e^{2i\delta(p,\ell,S,j)}}{2ip} \,.
\label{Eq:tmatrix}
\end{equation}
Here $\ell, S$, and $j$ are the relative orbital angular momentum,
the neutron spin $S=1/2$, and the total angular momentum,
respectively, and $\delta(p,\ell,S,j)$ are the phase shifts,  while $\mred$ is the
reduced mass for the $n\alpha$ system.

 Indeed,  we find that the resonance leads to an
enhanced contribution in the neutron spin structure function at temperatures
in the range of $0.1-4$~MeV. A comparison with the $nn$ scattering case is shown in Fig.~\ref{emiss}. Specifically, the energy loss $Q$ is shown as a function of the temperature $T$.  $Q$  is divided by $T^5$ (in MeV), by the neutron mass
fraction $f_n$, and by the mass density over the mass number $\rho_{X,12/A_X}$ of
$X = n$ or $\alpha$ particles (in $10^{12}$ g cm$^{−3}$), with $A_X
= 1$ or 4, respectively.
 We observe that, for the same
neutron mass fraction $f_n$ and if the $\alpha$-particle density
$\rho_\alpha/4 = m_N \, n_\alpha$ is comparable to $\rho_n = m_N \,
n_n$, $n\alpha$ bremsstrahlung dominates over $nn$ for $T <
6$~MeV. 

We have included the $D$-wave contribution in our calculations
and find that for the temperatures of interest they can be neglected.
Finally, even though
  for energies greater than $20$~MeV 
$n\alpha$ scattering has inelastic channels,  which can be
parameterized by an imaginary part in the phase shifts, the Boltzmann factors are small at such large energy, making their effects negligible.

\section{Conclusions}
We have presented results for bremsstrahlung rates for $\nu\bar{\nu}$ production in
$nn$ scattering processes for degenerate neutrons~\cite{Bacca09}, and
$nn$~\cite{Bacca12} and $n\alpha$~\cite{Sharma15} scattering processes when neutrons  and $\alpha$-particles are
nondegenerate.
For the $nn$ case, the used chiral N$^3$LO interactions generally
decreases the bremsstrahlung rates compared to the  one-pion-exchange potential by a factor of about $2-3$.

The $P$-wave resonance for $n\alpha$ scattering near $1$~MeV center of mass
energy leads to an enhancement in the bremsstrahlung rate for $T<6$~MeV. For
$T\lesssim2$~MeV, for $\alpha$ number densities comparable to neutron number
densities, the $n\alpha$ constitution is over an order of magnitude larger
than the contribution from $nn$ bremsstrahlung. Since these temperatures are
relevant for proto-neutron stars, it would be interesting to check whether this
enhanced rate affects the final predicted spectrum of neutrinos and/or the
dynamics of the dense matter outside the proto-neutron star. This can be
accomplished by including the $n\alpha$ process in the simulations of
core-collapse supernovae and proto-neutron start cooling.

\vspace{1cm}
{\bf Acknowledgments.} This work was supported in parts by the Natural Sciences and Engineering Research Council (NSERC), the National Research Council of Canada, and the ERC Grant No.~307986.

%%%%%%%%%%%%%%%%%%%%%%%%%%%%%%%%%%%%%%%%%%%%%%%%%%%%%%%%
% End of the paper
%

\end{document}